\documentclass{PoS}

\title{On the tension between Large Scale Structures and Cosmic Microwave Background}

\ShortTitle{Tension between LSS and CMB}

\author{\speaker{Marian Douspis}\\
        Institut d'Astrophysique Spatiale, CNRS, Université Paris Sud, Orsay, France\\
        E-mail: \email{marian.douspis@ias.u-psud.fr}}

\author{Laura Salvati\\
        Institut d'Astrophysique Spatiale, CNRS, Université Paris Sud, Orsay, France\\
        E-mail: \email{laura.salvati@ias.u-psud.fr}}

\author{Nabila Aghanim\\
        Institut d'Astrophysique Spatiale, CNRS, université Paris Sud, Orsay, France\\
        E-mail: \email{nabila.aghanim@ias.u-psud.fr}}

\abstract{Recent years have brought strong observational evidences for
  the standard LCDM cosmological model. Cosmic microwave background
  (CMB) anisotropy and large scale structure (LSS) probes do not
  favour any extensions of the standard model. Nevertheless, in this
  framework, the prefered cosmological parameters may differ from
  probe to probe, from experiment to experiment. This is the well
  known case of the tension between CMB and Sunyaev Zel'dovich (SZ)
  galaxy clusters (GC) from Planck\footnote{planck.esa.int}. In 2013,
  the Planck team has shown that the prefered matter content
  ($\Omega_M$) and density fluctuation power spectrum amplitude
  ($\sigma_8$), the two main cosmological parameters probed by the
  galaxy cluster number count, are different in the CMB analyses and
  in the SZ cluster analyses at more than 2 sigmas (a result confirmed
  in subsequent analyses). We present the results of our new analysis
  using more recent measurements of the CMB, SZ clusters and SZ power
  spectrum of 2016 and show that the tension on
  ($\Omega_M$,$\sigma_8$) is mostly releaved. The lower value of the
  reionisation optical depth and thus of $\sigma_8$ in the recent
  Planck studies is the main reason. We also show that basic
  extensions of the standard model (massive neutrinos or non--lambda
  dark energy) do not help improving the agreement between the
  probes. In order to fully reconcile SZ clusters with CMB best model,
  the mass of the galaxy clusters derived from hydrostatic equilibrium
  should be 40\% lower than the true mass. While current numerical
  simulations and weak lensing measurements agree for a mass bias of
  20\%, investigations are still going on to explain such disagreement
  on the mass bias. We show that considering a mass bias evolving with
  redshift or mass does not help in eliminating the discrepancy.}

\FullConference{2nd World Summit: Exploring the Dark Side of the Universe\\
		25-29 June, 2018\\
		University of Antilles, Pointe-à-Pitre, Guadeloupe, France}

\begin{document}

\section{A cosmological tension ?}

\begin{figure}
     \includegraphics[width=.5\textwidth]{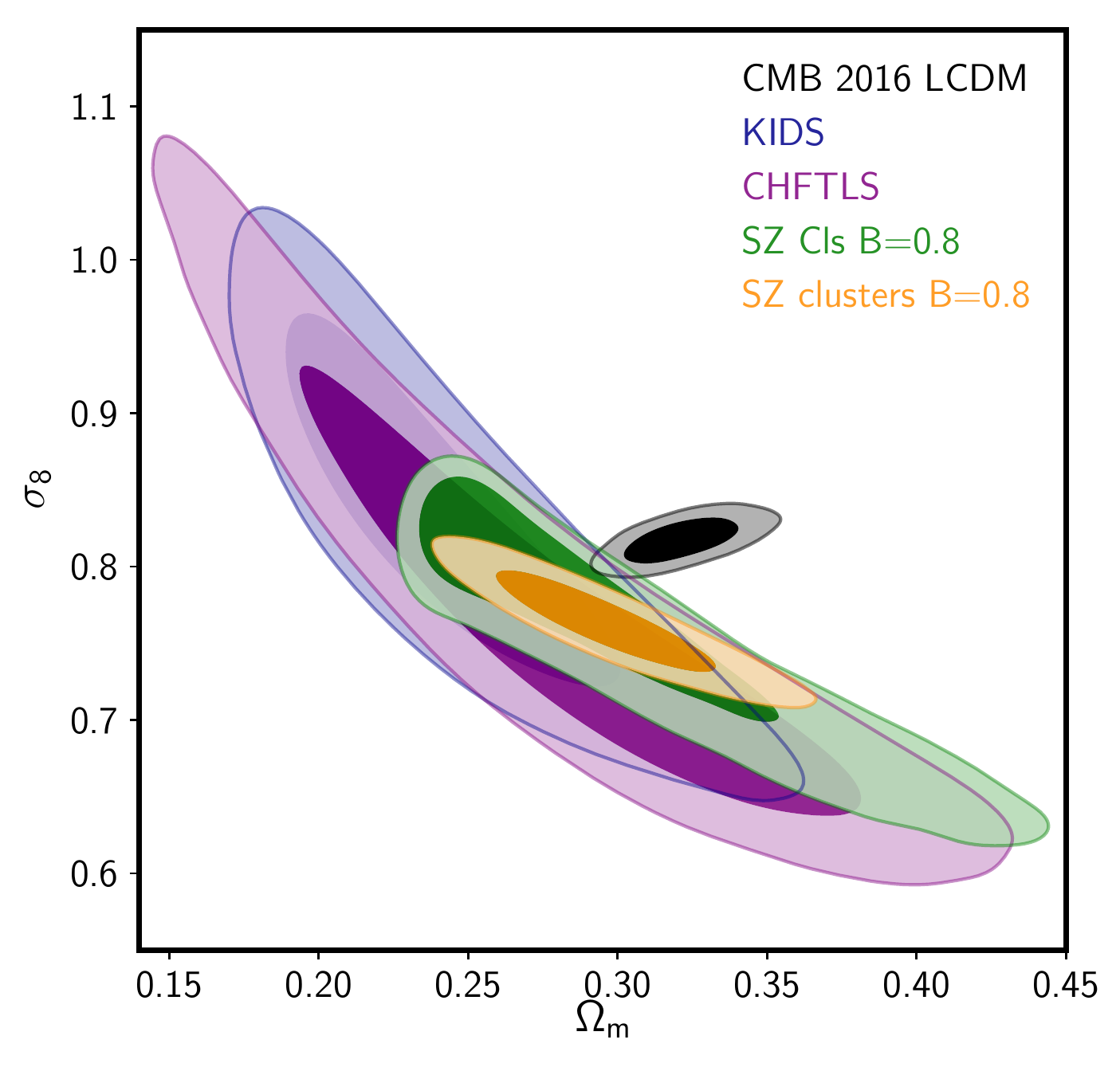}\includegraphics[width=.5\textwidth]{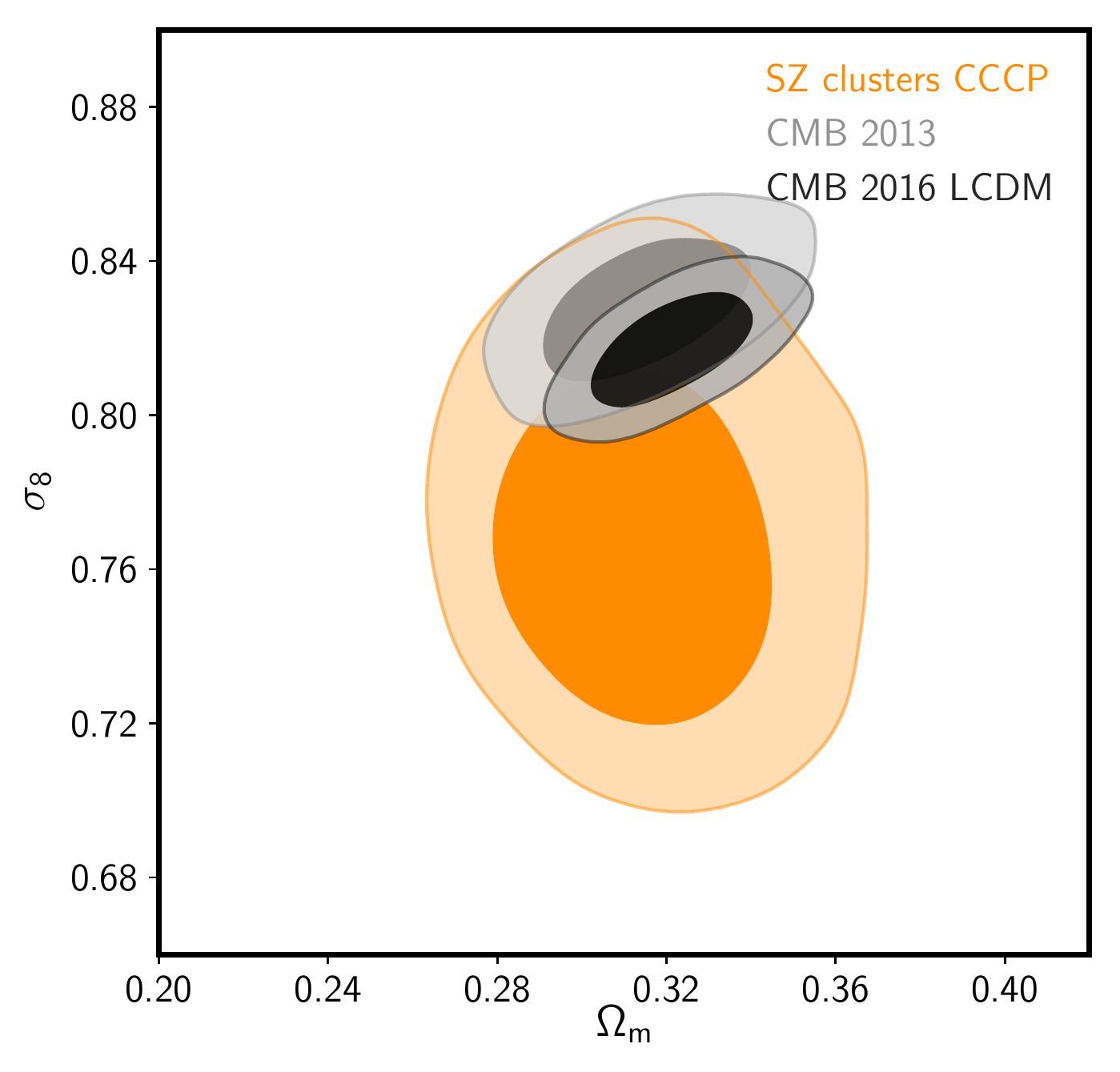}
     \caption{Left: recent large scale constraints compared with CMB
       (black filled contours). See text for references. Right:
       Comparison of SZ cluster constraints from Planck (yellow) with
       CCCP prior on the mass bias and Planck CMB constraints from
       2013 and 2016 (grey and black respectively). The displacement
       of the CMB constraints is due to the new estimate of the
       reionisation optical depth. Adapted from \cite{SDA18}.}
     \label{fig:OS}
\end{figure}

Galaxy clusters are thought to be strong probes of the cosmological
model because of their large scales, dark matter dominated content and
thus sensitivity to the global evolution of the Universe rather than
baryonic micro-physics. Samples of GC have been used in several
frequency domains (optical, X-ray, SZ) to put constrains on
cosmological parameters. Only recently, samples
were large enough to make results almost insensitive to statistical
errors. This has been the case in SZ, when Planck published it first
catalogue \cite{PSZ} and used a subsample of circa $200$ clusters
\cite{P13} to probe the standard cosmological model. Constraints form
GC were shown to be 2.4-$\sigma$ away from the CMB (Fig.~\ref{fig:OS},
right). In parallel, in the last years, galaxy surveys (e.g. KIDS,
CFHTLS, DES) have produced weak lensing constraints on the same
parameters showing as well some tension with CMB results. While all
experiments contain systematics that may explain the small shift in
prefered cosmological parameters, it is puzzling that all low redshift
large scale structure probes are in agreement and drive $\sigma_8$
towards lower values than the CMB (Fig.~\ref{fig:OS}, left).

In \cite{SDA18} we have performed a new analysis of the SZ signal from
Planck, combining the number counts in redshift and signal-to-noise
ratio (NC), of circa 400 SZ clusters \cite{P15SZ}, with the SZ power
spectrum (CL) derived from 50\% of the MILCA SZ map
\cite{MILCA,P15CL}. We compared and combined the derived constraints
with the CMB assuming the new value of the optical depth of
reionisation $\tau=0.055\pm 0.009$ from Planck \cite{P16tau}.

\begin{figure}
     \includegraphics[width=.5\textwidth]{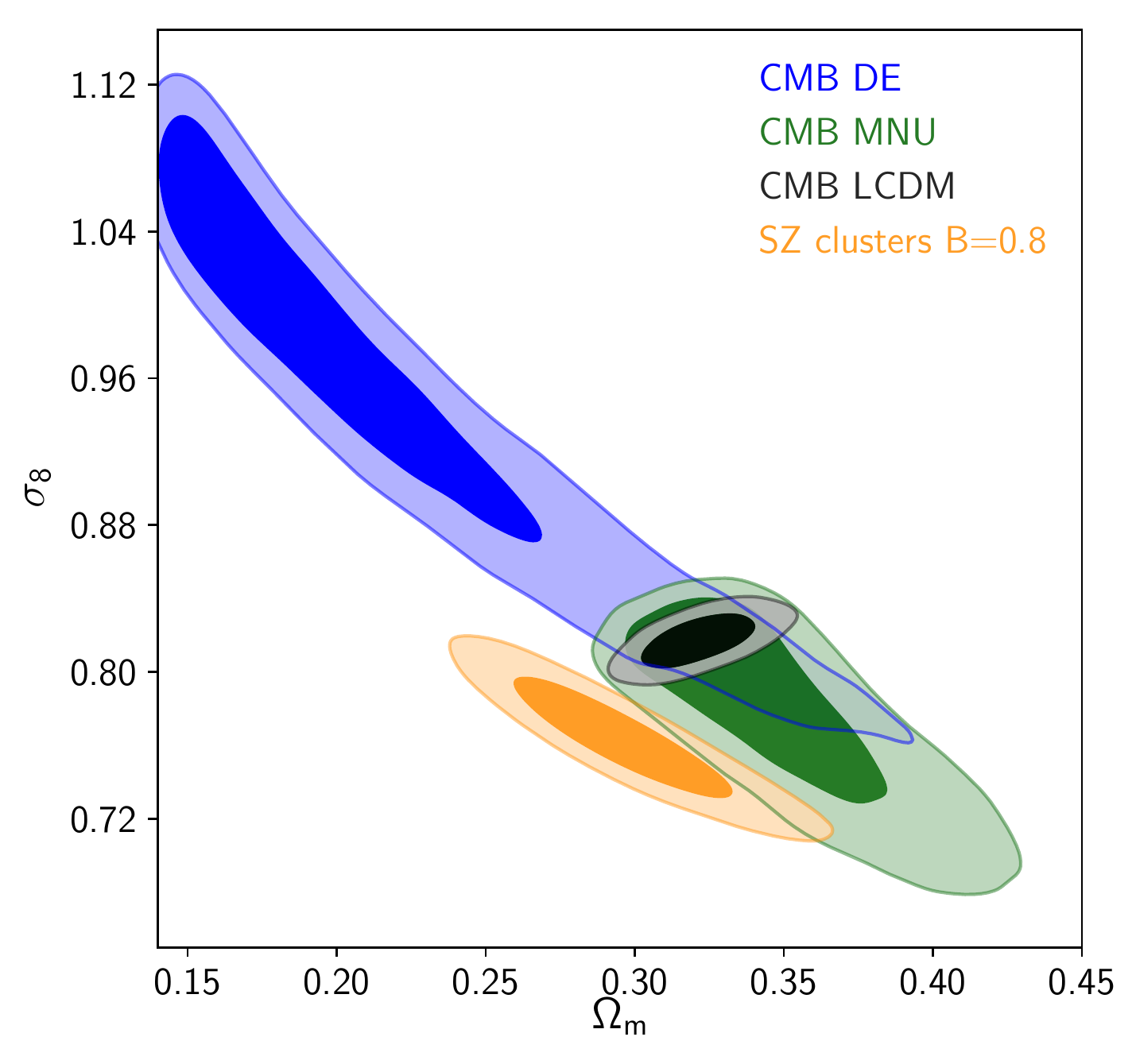}\includegraphics[width=.5\textwidth]{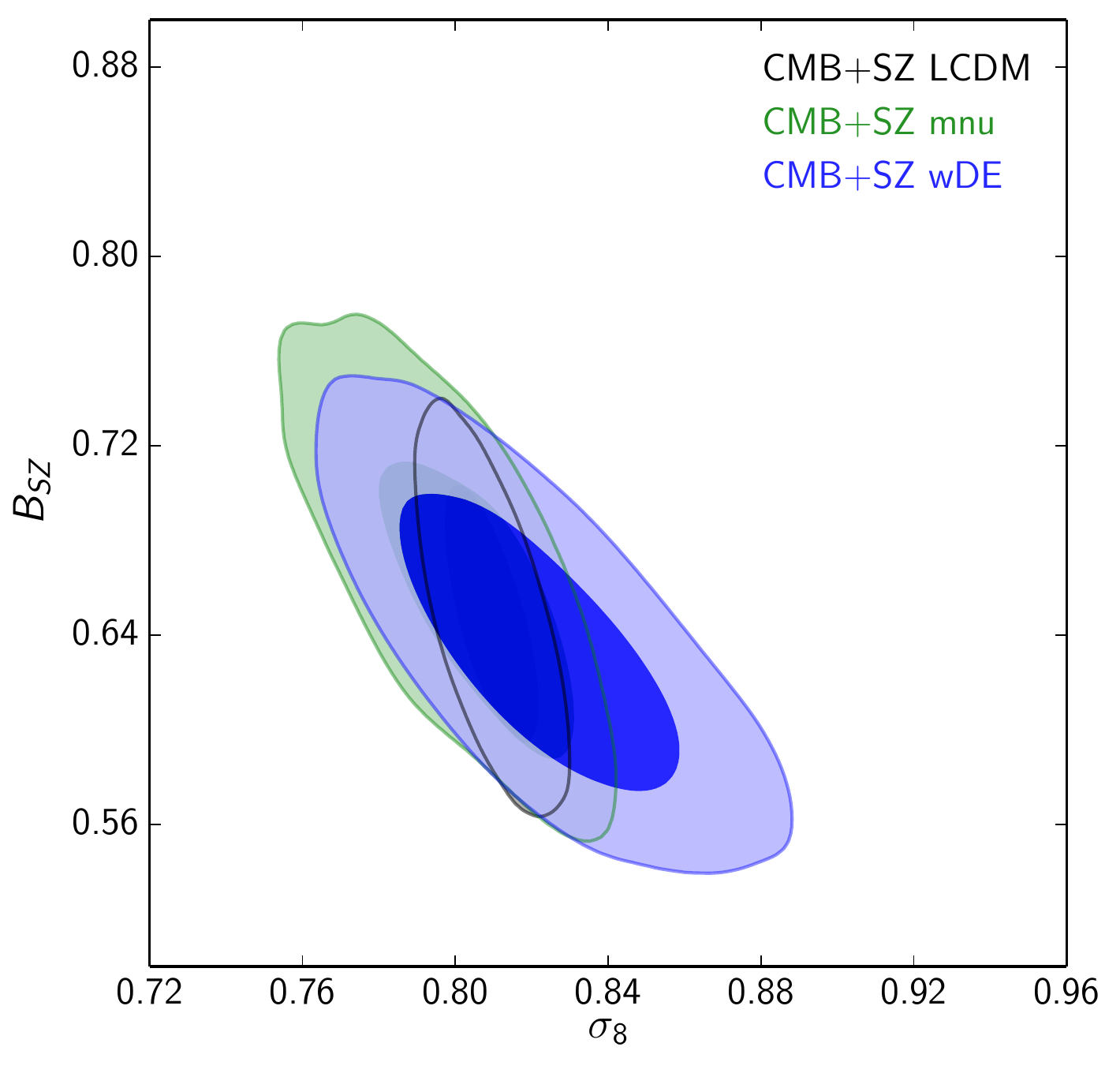}
     \caption{Left: Comparison of cosmological paramater when opening
       up the extensions of LCDM (massive neutrinos (green) and dark
       energy (blue)) for CMB. LCDM contours are in black. Right:
       Constraints on the mass bias ($B_{SZ}$) and $\sigma_8$ for the
       different cosmological model extensions when combining CMB and
       SZ.  }
     \label{fig:OSExt}
\end{figure}

We show in Fig.~\ref{fig:OS} (right) that the tension is reduced to
1.5 $\sigma$ with such a new value of $\tau$.  A full agreement
between the two probes is not yet completely achieved when assuming
the default value of the mass bias, $(1-b)\sim 0.8$ (CCCP prior
\cite{CCCP}, see next section), and new tests on extensions of the
standard cosmological model do not help (Fig.~\ref{fig:OSExt},
left). In the three cosmological models we probe (LCDM, massive
neutrinos, dark energy), the prefered value of the mass bias is always
much lower than the default value ($\sim 0.8$) as shown in
Fig.~\ref{fig:OSExt} (right). A value of $(1-b)\sim0.62$ (a 38\% lower
hydrostatic estimate compared to the true mass) is needed to reconcile
CMB and SZ probes. Furthermore, such a value leads to a baryon fraction
in clusters at odd with the universal value \cite{Eckert}.

\section{An astrophysical tension ?}

\begin{figure}
  \center
     \includegraphics[width=.6\textwidth]{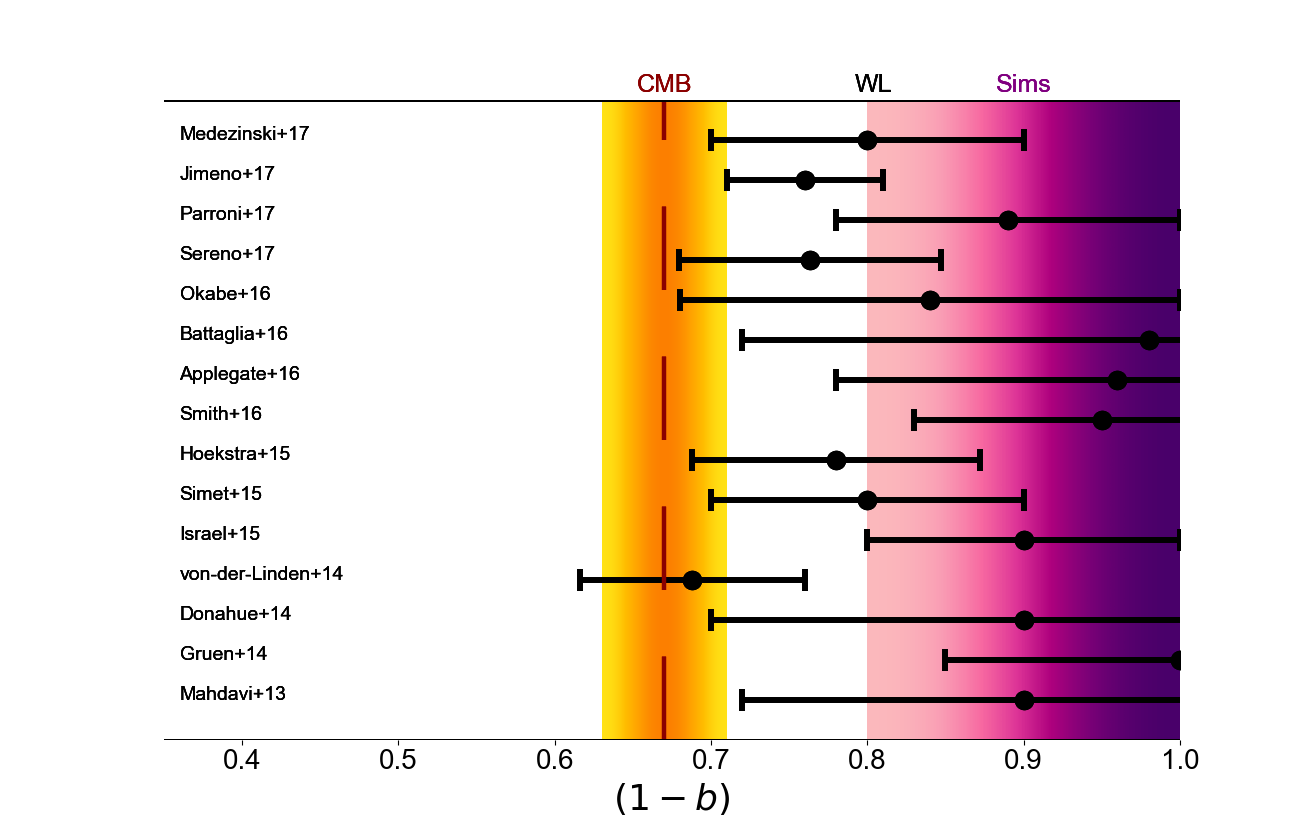}
     \caption{Mass bias estimates from simulations (purple dashed area) and lensing estimates from different teams (black points with errors), compared the value needed for reconciling CMB and SZ (orange area).}
     \label{fig:bias}
\end{figure}

The main systematic of the SZ cluster analyses is the mass bias: the
ratio between the hydrostatic mass estimates (HME) and the true
mass. The HME are derived from X-ray and SZ data \cite{PSZ}. The true
mass is not an observable and can be assumed to be derived from
numerical simulations (NS) or lensing measurements (lensing mass
estimates, LME). In \cite{P13} the bias (1-b) is computed by comparing
the observed mass (from Xray and SZ) with a set of 12 numerical
simulations (see citations therein). In the following studies, a prior
on the bias coming from HME and LME comparison is usually
assumed. Figure \ref{fig:bias} summarizes the recent values of the
bias derived from both NS and LME. As stated above, the low value of
$(1-b)$ needed to retrieve CMB cosmology from clusters is marginally
consistent with other estimates. A unique constant value of the mass
bias cannot thus explain the remaining tension between CMB and SZ.

\begin{figure}
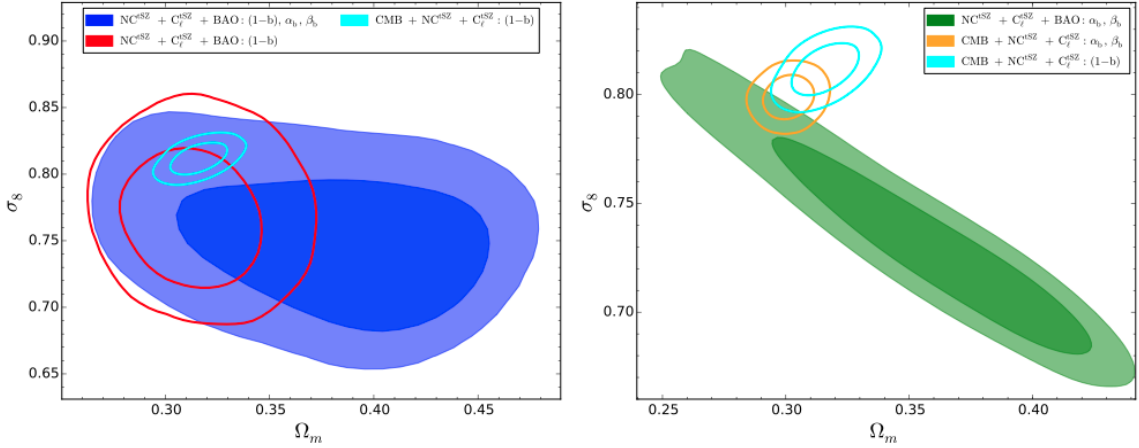

     \includegraphics[width=.5\textwidth]{Figs/OS_bab_SZ.pdf}\includegraphics[width=.5\textwidth]{Figs/OS_bab.pdf}
     \caption{Constraints from different test of SZ and CMB
       probes. Left: default constraints from SZ (red) and CMB
       (cyan) are compared with SZ constraints with evolving bias
       free (blue). Right: the light green area shows the SZ
       constraints assuming a fixed amplitude of the bias to 0.75 but
       letting the mass and redshift indeces free. The yellow contours
        is given by the combination of the latter with CMB, and
       offers a really bad goodness of fit of both CMB and SZ data. Adapted from \cite{SDA18+}.}
     \label{fig:evolB}
\end{figure}

In order to investigate further the solutions to reconcile the 2
probes, we study the possibility that the mass bias is in fact
depending on the mass and redshift of the clusters. If the HME is not
a good proxy for the total mass, it may be due to baryonic effects,
non thermal pressure, magnetic field, ...; processes that may be non
universal and linear, and vary with the mass and redshift of the
clusters. We thus consider the following evolution of the mass bias
\cite{SDA18+}:
\begin{equation}
(1-b)_{\rm var} = (1-b) \cdot \left( \frac{M}{M_*}\right)^{\alpha}\cdot \left( \frac{1+z}{1+z_*}\right)^{\beta} \, ,
\end{equation}
where $(1-b)$ is an amplitude at a given mass and redshift, $M_*=6
\cdot 10^{14} M_{\odot}$ to be consistent with the pivot mass of the
scaling relations (see Eqs. (7) and (8) in \cite{P15SZ}) and
$z_*=0.22$ is the median value of the clusters catalog that we are
considering.  When all bias parameters ($1-b, \alpha, \beta$) are let
free in addition to the cosmological parameters, the contraints of SZ
obviously weakens but are not in better agreement with CMB ones as
shown in Fig.~\ref{fig:evolB} (left). In addition, we don't find
evidence for any evolution of the bias with mass and redshift at more
than 1 $\sigma$. We reach the same conclusions when combining CMB and
SZ probes. Furthermore, when the mass bias is fixed to an optimistic
value of $(1-b)=0.75$ at the mean mass and redshift of the cluster
sample, allowing for redshift and mass dependencies does not drive the
cluster constraints towards the CMB ones. Figure \ref{fig:evolB}
(right) shows the contours of SZ alone with such a fixed bias (green), compared to CMB (cyan). The yellow constraints are
obtained when combining CMB and SZ with fixed $(1-b)$ and free mass
and resdshift indeces. The goodness of fit of such combination is very
bad, showing that no reasonable evolving bias exists to reconcile CMB
and SZ probes.

\section{Conclusions}

The tension between the cosmological models prefered by CMB and SZ
data is reduced thanks mostly to the new value of the optical depth of
reionisation observed by Planck since 2016, but both probes lead to
differents views on the mass estimates of clusters. Simple dark energy
models (constant equation of state) and massive neutrinos are not
appropriate extensions to reduce fully the remaining tension. Modified
gravity scenari may provide a good cosmological model allowing to
satisfy both probes, but in simple scenarios, extreme departure from
general relativity are needed \cite{Ziad}. In the LCDM model, the
``simple'' way to reconcile both CMB and SZ is to require a high mass
bias (low value of $(1-b)\sim0.6$), implying 40\% missing mass when
estimated through hydrostatic equilibrium. Such a low value is not
supported by numerical simulations nor lensing mass measurements. When
this bias is allowed to be evolving with redshift and mass, the
cluster constraints weakens but do not reach the CMB
ones. Furthermore, no evidence for varying bias is found. Being it
constant or evolving, no value of the bias is found to satisfy for SZ
clusters and CMB observations.

\acknowledgments{} Based on observations obtained with Planck
(http://www.esa.int/Planck), an ESA science mission with instruments
and contributions directly funded by ESA Member States, NASA, and
Canada. Part of this research has been supported by the funding
for the ByoPiC project from the European Research Council (ERC) under
the European Union's Horizon 2020 research and innovation programme
grant agreement ERC-2015-AdG 695561.

\end{document}